\documentclass[conference]{IEEEtran}
\IEEEoverridecommandlockouts
\usepackage{cite}
\usepackage{amsmath,amssymb,amsfonts}
\usepackage{algorithmic}
\usepackage{graphicx}
\usepackage{textcomp}
\usepackage{xcolor}
\usepackage{fancyhdr}
\usepackage{subcaption}
\usepackage{multirow}
\usepackage[inline]{enumitem}
\usepackage{gensymb}

\definecolor{yellow}{rgb}{1,1,0}
\definecolor{darkblue}{rgb}{0.4,0.7,1}
\definecolor{lightblue}{rgb}{0.7,0.9,1}

\def\BibTeX{{\rm B\kern-.05em{\sc i\kern-.025em b}\kern-.08em
    T\kern-.1667em\lower.7ex\hbox{E}\kern-.125emX}}

\fancypagestyle{firstpage}{
  \fancyhf{}

  \fancyhead[C]{\normalsize{To appear at ISVLSI 2021}} 
}
   
\begin{document}

\title{A Microarchitecture Implementation Framework for Online Learning with Temporal Neural Networks
}

\author{\IEEEauthorblockN{Harideep Nair}
\IEEEauthorblockA{\textit{Electrical and Computer Engineering} \\
\textit{Carnegie Mellon University}\\
harideep.nair@sv.cmu.edu}
\and
\IEEEauthorblockN{John Paul Shen}
\IEEEauthorblockA{\textit{Electrical and Computer Engineering} \\
\textit{Carnegie Mellon University}\\
jpshen@cmu.edu}
\and
\IEEEauthorblockN{James E. Smith}
\IEEEauthorblockA{\textit{Electrical and Computer Engineering} \\
\textit{University of Wisconsin (Emeritus)}\\
\textit{Carnegie Mellon University (Adjunct)}\\
jes@ece.wisc.edu}
}
\maketitle
\thispagestyle{firstpage}

\begin{abstract}
Temporal Neural Networks (TNNs) are spiking neural networks that use time as a resource to represent and process information, similar to the mammalian neocortex.
In contrast to compute-intensive deep neural networks that employ separate training and inference phases, TNNs are capable of extremely efficient online incremental/continual learning and are excellent candidates for building edge-native sensory processing units.
This work proposes a microarchitecture framework for implementing TNNs using standard CMOS. Gate-level implementations of three key building blocks are presented: 1) multi-synapse \textit{neurons}, 2) multi-neuron \textit{columns}, and 3) unsupervised and supervised online learning algorithms based on \textit{Spike Timing Dependent Plasticity (STDP)}. 
The proposed microarchitecture is embodied in a set of characteristic scaling equations for assessing the gate count, area, delay and power for any TNN design. Post-synthesis results (in 45nm CMOS) for the proposed designs are presented, and their online incremental learning capability is demonstrated.
\end{abstract}


\section{Introduction}

Current computing demand for training Deep Neural Networks (DNNs) is 
doubling every 3.4 months \cite{openai}. Moore's law, at best, is only doubling every 2 years. The gap between increasing computing demand and what computing hardware can provide is widening at the rate of 8x per year. 
This calls for new paradigms and new types of hardware that are orders of magnitude more efficient for performing human-like sensory processing and online learning \cite{schwartz2019green}. Neuromorphic temporal neural networks appear to exhibit such potential.

\begin{figure}[t]
    \centering
    \includegraphics[width=3.5in]{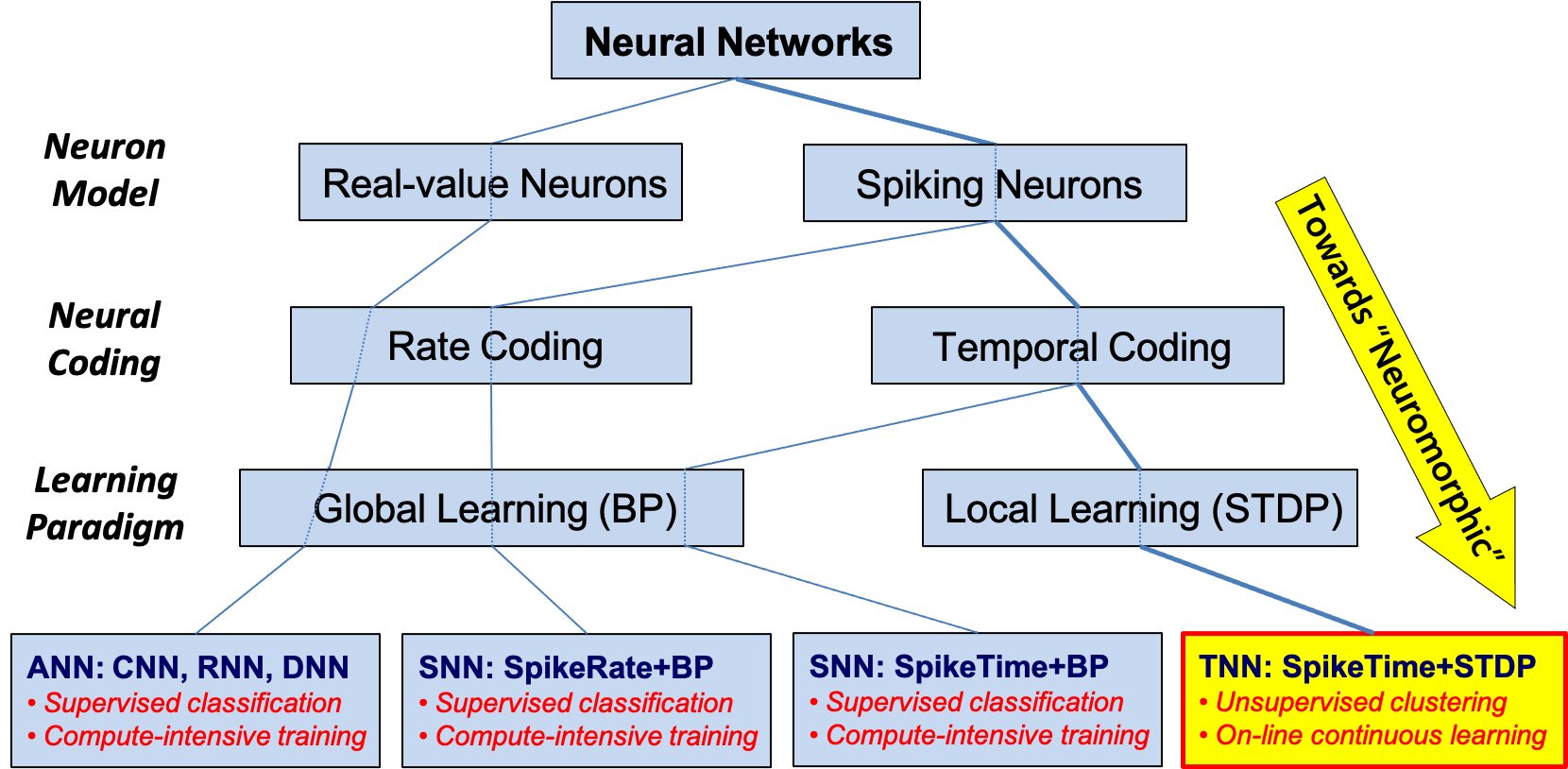}
    \caption{Neural Network Taxonomy}
    \label{tax}
\end{figure}

Temporal Neural Networks (TNNs) \cite{smith2018space,smith2020neuromorphic,smith2020temporal} strive to achieve not just the behavior/function of biological neural networks but also their structure/organization. TNNs adhere to biological plausibility with the goal of achieving brain-like capability and efficiency. Fig. \ref{tax} highlights the distinctive neuromorphic attributes of TNNs.  TNN components communicate via spikes like all Spiking Neural Networks (SNNs) \cite{maass1997networks,natschlager1998spatial}. However, TNNs belong to a special class of SNNs that encode and process information in temporal form using precise \textit{spike time relationships}, unlike most SNNs that use \textit{spike rates} \cite{kim2020spiking, diehl2015unsupervised, sengupta2019going}
for information encoding and processing. TNNs also employ a form of \textit{local} learning called Spike Timing Dependent Plasticity (STDP) \cite{guyonneau2005neurons}, as opposed to \textit{global} backpropagation and stochastic gradient descent commonly used in DNNs \cite{lecun2015deep} and SNNs \cite{bohte2002error, mostafa2017supervised, lee2016training}.

TNNs fueled by STDP are capable of learning in an online, incremental, continual fashion \cite{smith2020neuromorphic, smith2020temporal} and therefore provide a promising technology for building sensory processing units in mobile/edge devices. The efficacy of TNNs in performing unsupervised time-series clustering for various edge-native applications such as anomaly detection, healthcare monitoring, etc. has been shown in \cite{chaudhari2021unsupervised}. This work builds on recent work in \cite{smith2018space} which lays the foundation of TNNs as space-time computing networks based on a rigorous space-time algebra. In \cite{smith2020neuromorphic,smith2020temporal} it is proposed that one can implement a \textit{silicon neocortex} capable of brain-like online learning
by examining the organization of biological neural networks to formulate an analogous architecture for TNNs. 
We follow this approach by focusing on direct hardware implementation of TNNs.

This work explores the practical feasibility of direct hardware implementation of TNNs using standard digital CMOS technology. In a direct implementation, hardware clock cycle is used as the basic time unit for temporal processing, i.e., time itself is not stored as a binary value but implicit in the clock. 
We define a TNN microarchitecture and implement its key building blocks: 1) multi-synapse \textit{neurons}, 2) multi-neuron \textit{columns} and 3) \textit{STDP} (unsupervised) and \textit{R-STDP} (supervised with reward) online learning algorithms. We present their gate-level designs along with characteristic scaling equations for estimating the area, power  and delay for any arbitrary TNN. A distinct feature of the proposed framework is a novel synapse design that integrates weight storage with synaptic processing, thereby eliminating the need for a separate weight storage. 
To the best of our knowledge, this is the first work that presents a microarchitecture framework for directly implementing TNNs capable of online learning.

\section{TNN Organization and Operation}

\subsection{Temporal Encoding and Processing}
\begin{figure}[t]
    \centering
    \includegraphics[width=3.16in]{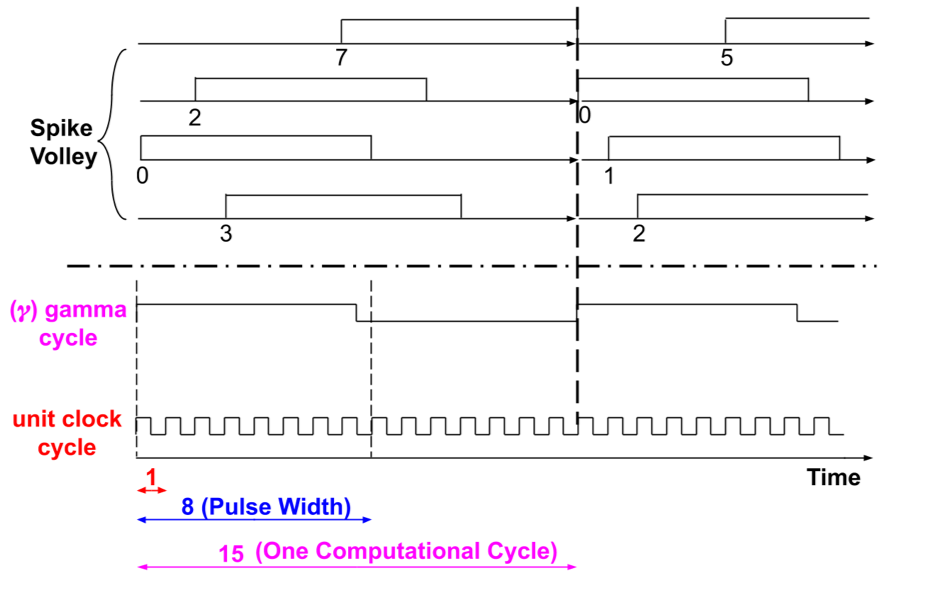}
    \caption{Temporal Encoding and Processing}
    \label{tcoding}
\end{figure}

A distinctive attribute of TNNs involves the use of temporal encoding, wherein information is represented by relative timings of spikes. In a TNN, computation occurs in volleys or waves of spikes. A volley consists of at most one spike per synaptic input. 
As shown in Fig. \ref{tcoding}, two clocks are used. The \textit{unit clock} is the finest temporal resolution and is also the synchronizing clock used in the digital hardware. The \textit{gamma clock} (inspired from the biological gamma cycles \cite{fries2007gamma}) frames the computing window and is the time required for a column to communicate and process a spike volley and update synaptic weights.
This implementation studied here uses 3 bits of precision for temporal encoding and synaptic weights.
Spikes in a volley are implemented as unit time pulses, a form of unary encoding, and volleys are separated using gamma clock cycles. With unary encoding, it takes up to 7 time units to encode a 3-bit value. Allowing additional time to process a spike volley, the gamma cycle is extended to 15 time units. This is explained in further detail in Section \ref{fsmsyn_sec}.

\subsection{Key TNN Building Blocks}

%

\begin{figure}[t]
    \centering
    \begin{subfigure}[b]{0.53\columnwidth}
        \centering
        \includegraphics[width=\columnwidth]{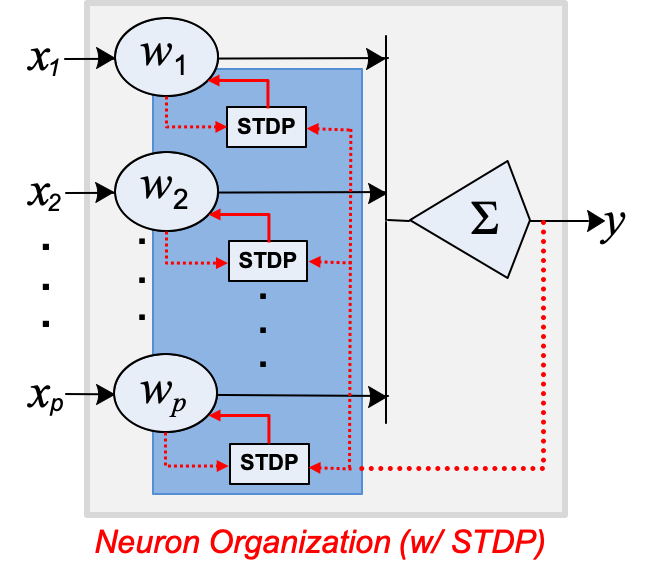}
        \caption{Neuron: \textit{p} Synapses, STDP}
        \label{fig3a_new}
    \end{subfigure}%
    \begin{subfigure}[b]{0.24\textwidth}
        \centering
        \includegraphics[width=\columnwidth]{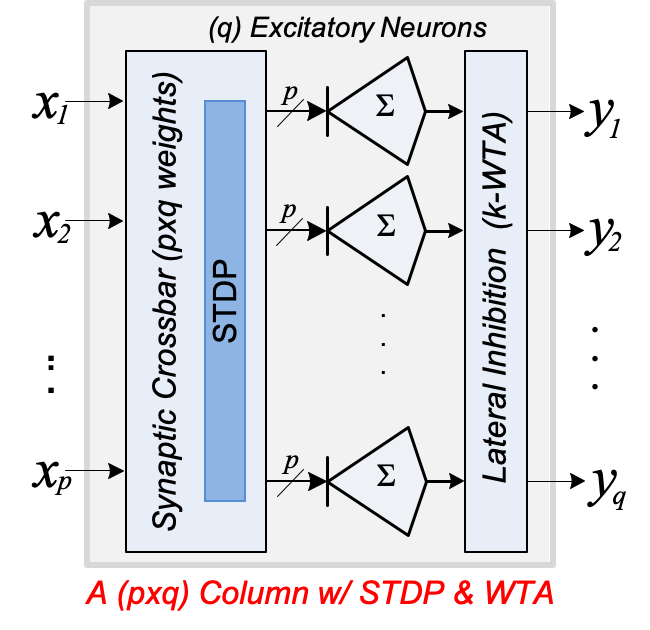}
        \caption{Column: q Neurons \& WTA}
        \label{fig3b_new}
    \end{subfigure}
    \caption{Key TNN Building Blocks}
\end{figure}

The most fundamental TNN building block is a neuron. As shown in Fig. \ref{fig3a_new}, each neuron has \textit{p} synaptic inputs and one output. Each synaptic input carries a synaptic weight, which is updated locally based on the relative timing of the incoming spike to that synapse and the outgoing spike from the associated neuron body. The rules for updating synaptic weights constitute the STDP learning method - the key building block that imparts TNNs their functionality. Through STDP, a neuron learns an input feature by adapting its synaptic weights to closely match the corresponding input pattern.

The smallest operational building block is a column which, in itself, is a fully-functional TNN. As shown in Fig. \ref{fig3b_new}, a column is a stack of \textit{q} parallel neurons. Every neuron in a column shares the same set of \textit{p} inputs, known as a \textit{receptive field}. A \textit{p$\times$q} synaptic crossbar contains \textit{p$\times$q} synaptic weights, each of which is updated by STDP. On the output side of the \textit{q} neurons, one winner-take-all (1-WTA) \textit{lateral inhibition} is performed by selecting the earliest spiking neuron from among the \textit{q} neurons as the one winner. Output spiking is disabled for non-winning neurons.
This introduces competition among the neurons and enables the column to learn a set of distinct features local to its input receptive field.

This paper presents the CMOS implementation of a neuron (Section \ref{neuron_sec}) and a column (Section \ref{column_sec}). In Section \ref{stdp_sec}, STDP rules for updating synaptic weights are discussed. The baseline STDP method is unsupervised. We also introduce a variation, called \textit{reinforcement} STDP, which is similar to the \textit{reward modulated} STDP in \cite{mozafari2019bio}. Post-synthesis and online learning evaluations are performed in Sections \ref{uarch_eval} and \ref{online_eval} respectively.

\section{Neuron Implementation}
\label{neuron_sec}
This work adopts the widely used Spike Response Model \cite{kistler1997reduction} SRM0. This section presents the components of this excitatory neuron model and their detailed gate level designs.

\subsection{Synaptic Response Functions}
A \textit{synapse} connects the \textit{axon} (output) of a pre-synaptic neuron and a \textit{dendrite} (input) of the post-synaptic neuron. An SRM0 neuron takes multiple input spikes and generates a response function for each spike based on its corresponding synaptic weight. All the individual response functions are then added to form the neuron’s membrane potential. When (and if) the membrane potential crosses a threshold, the neuron fires an output spike on its axon.  
%
In this work, we adopt the ramp-no-leak (RNL) function for its temporal computational benefits and implementation efficiency \cite{smith2020neuromorphic, maass1997networks}. The RNL function increases by a unit step at every time unit until it reaches its peak and then remains constant until it is reset prior to the next computation cycle. The ``ramp'' allows responses from different synapses to be distributed temporally based on the synaptic strengths (weights), which proves to be particularly powerful for TNNs that operate temporally. The no-leak model is based on arguments that the leak is primarily a reset mechanism \cite{guyonneau2005neurons, masquelier2007unsupervised}. For silicon implementations, there are simpler ways to reset at the end of each gamma cycle.

\subsection{Synapse Modeling}
\label{fsmsyn_sec}
\begin{figure}[t]
        \centering
        \includegraphics[width=3.1in]{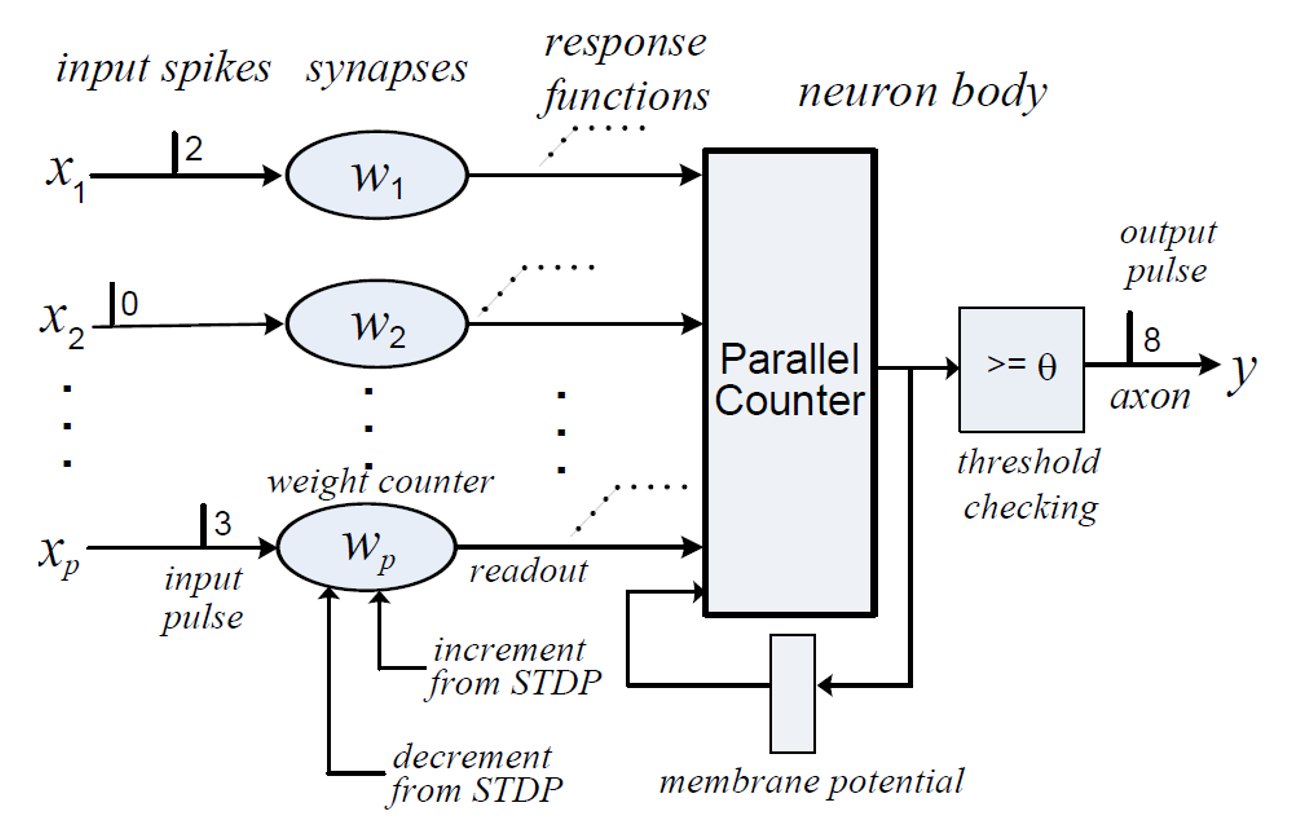}
        \caption{SRM0 Neuron with RNL Response Function}
        \label{pt}
\end{figure}
Fig. \ref{pt} shows the block diagram for the proposed SRM0 neuron implementing RNL response function. Its operation consists of three main stages: 1) temporal arrival of input spikes, 2) serial thermometer readout of RNL response functions based on the corresponding synaptic weights, and 3) binary accumulation of thermometer-coded response functions into the membrane potential.
%
Synapses are implemented as finite state machines (FSMs) operating as binary counters. If the maximum weight is $w_{max}$, the number of counter bits is $ceiling(log_2(w_{max} + 1))$. The counter has three modes, two controlled by STDP (described in Section \ref{stdp_sec}): increment (up to $w_{max}$) and decrement (down to 0). The third \textit{readout} mode is controlled by the input pulse.
Readout mechanism is meticulously integrated into the same FSM used for storing synaptic weight and is described below.

As will become apparent, synapses dominate hardware complexity and hence the synapse design must be highly optimized. Our approach uses a pulse of width $w_{max}+1$. The input pulse directly controls the counter readout. When the leading edge of an input pulse occurs (0$\rightarrow$1 transition), the weight counter is decremented and an output of 1 is emitted each unit clock cycle until the counter reaches 0. This essentially converts the binary weight value in the counter to a serial thermometer code. 
After the counter reaches 0, it wraps around to $w_{max}$ and continues to count down until the trailing edge of the input pulse (1$\rightarrow$0 transition) when
the weight in the counter is restored to its original value. Thus, once an input spike arrives, readout takes an additional 7 cycles. (Although we assume $w_{max}$ = 7 in this paper, this technique can be generalized to any $w_{max}$.)
STDP (Section \ref{stdp_sec}) takes another cycle. These coupled with 7 cycles for encoding give rise to a gamma period of 15 clock cycles.

In summary, a synapse and its weight are implemented with a counter FSM that can 1) increment, saturating at $w_{max}$; 2) decrement, saturating at 0; and 3) wrap-around decrementing, emitting an output of 1 prior to wrapping around and then a 0 thereafter. Note that this synapse design preserves the original weight value while doing RNL readout, which eliminates the need of a separate SRAM for weight storage and access.

\subsection{Neuron Body}
\begin{figure}[t]
        \centering
        \includegraphics[width=3.2in]{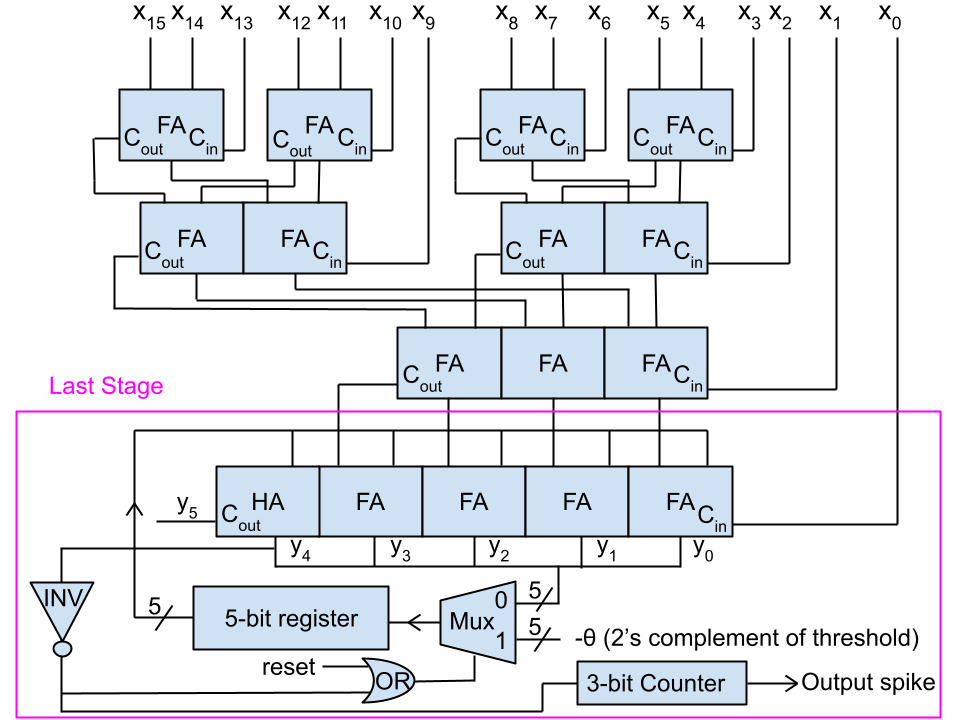}
        \caption{Neuron Body with 16 Synapses}
        \label{acc}
\end{figure}
The neuron body is implemented as a parallel counter that adds the thermometer coded weights coming from the synapses, cycle by cycle, thereby accumulating the membrane potential as a sum of RNL response functions. When (and if) the parallel counter output reaches the threshold $\theta$, an output spike is emitted during that cycle.

Based on Parhami \cite{parhami1995accumulative}, the membrane potential accumulator can be efficiently implemented using ripple carry adders by integrating a (\textit{p}-1)-input parallel combinational counter and a ($log_2p+1$)-bit adder into one design. Fig. \ref{acc} shows the design for a 16-input accumulator, with integrated output spike generation. For a \textit{p}-input accumulator, \textit{p}-1 inputs are accumulated into a ($log_2p$)-bit output, which is then added to the previous stored ($log_2p+1$)-bit value from the register with the one remaining input bit acting as carry-in. Note that the configuration in Fig. \ref{acc} allows all adder inputs to be efficiently utilized and is most optimal when \textit{p} is a power of 2. 

The accumulating register is initialized with (signed 2’s complement) -$\theta$ at every gamma cycle, which eliminates the need for any comparator for output spike generation. The  ($log_2p+1$)\textsuperscript{th} bit of the output indicates if the accumulated body potential has crossed the threshold and triggers a 3-bit counter to generates an 8-cycle wide pulse (output spike). 

\section{STDP \& R-STDP Implementation}
\label{stdp_sec}
STDP is a distinctive feature of TNNs. STDP learning is unsupervised and local to each synapse. It can perform inference and online continual learning at the same time. In this work, we  propose an STDP design that is both effective in learning and implementable using standard CMOS technology. 



\subsection{Proposed STDP Update Rules}
\begin{figure}
        \centering
        \includegraphics[width=2.9in]{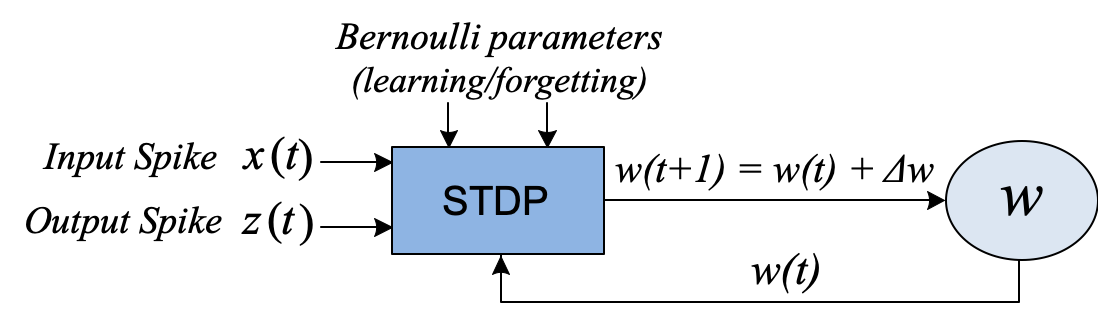}
        \caption{Local STDP Update Process}
        \label{stdp_illustrate}
\end{figure}

Our learning method is a customized version of the classic Spike Timing Dependent Plasticity (STDP) \cite{bi1998synaptic}. STDP is implemented locally at each synapse as shown in Fig. \ref{stdp_illustrate}. The proposed STDP learning rules are summarized in Table \ref{stdp}. Here, x(t) and z(t) represent input and output spiketimes respectively. $\Delta$w denotes change in weight and B($\mu$) represents a Bernoulli random variable with probability $\mu$.
\begin{table}[t]
  \centering
  \small
  \caption{STDP Update Rules}
  \scalebox{0.84}{
  \begin{tabular}{|c|c||c|}
    \hline
    \multicolumn{2}{|c||}{\textbf{Input Conditions}} & \textbf{Weight Update}\\
    \hline
    \hline
    $x(t)\neq \infty$; & $x(t)\leq z(t)$ & $\Delta w=+B(\mu_{capture})*max(F(w),B(\mu_{min}))$\\
    \cline{2-3}
    $z(t)\neq \infty$ & $x(t)>z(t)$ & $\Delta w=-B(\mu_{backoff})*max(F(w),B(\mu_{min}))$\\
    \hline
    \multicolumn{2}{|c||}{$x(t)\neq \infty$; $z(t)=\infty$} & $\Delta w=+B(\mu_{search})$\\
    \hline
    \multicolumn{2}{|c||}{$x(t)=\infty$; $z(t)\neq \infty$} & $\Delta w=-B(\mu_{backoff})*max(F(w),B(\mu_{min}))$\\
    \hline
    \multicolumn{2}{|c||}{$x(t)=\infty$; $z(t)=\infty$} & $\Delta w=0$\\
    \hline
  \end{tabular}
  }
  \label{stdp}
\end{table}

STDP update rules are divided into four major cases, corresponding to the four combinations of input and output spikes (represented by x(t) and z(t) respectively) being present ($\neq \infty$) or absent ($=\infty$) . When both are present, two sub-cases are formed based on the relative timing of the input and output spikes in the classical STDP manner \cite{bi1998synaptic}. In effect, a synaptic weight is incremented (strengthened) if there is an input spike and it either contributed (Case 1) or can potentially contribute (Case 3) to the output spike; else it is decremented.

The STDP update function either increments the weight by $\Delta$w (up to a maximum of $w_{max}$ = 7), decrements the weight by $\Delta$w (down to a minimum of 0), or leaves the weight un-changed. The $\Delta$ values (1, 0 or -1) are defined using Bernoulli random variables (BRVs) with parameterized learning probabilities denoted as B($\mu$) with a descriptive subscript. $F(w)$ is a stabilization function (=$B((w/w_{max})(1-w/w_{max}))$) which makes the weights "sticky" at both ends (0 and 7) \cite{kheradpisheh2016bio, kheradpisheh2018stdp}.

\subsection{Proposed STDP Implementation}
\begin{figure}[t]
        \centering
        \includegraphics[width=3.3in]{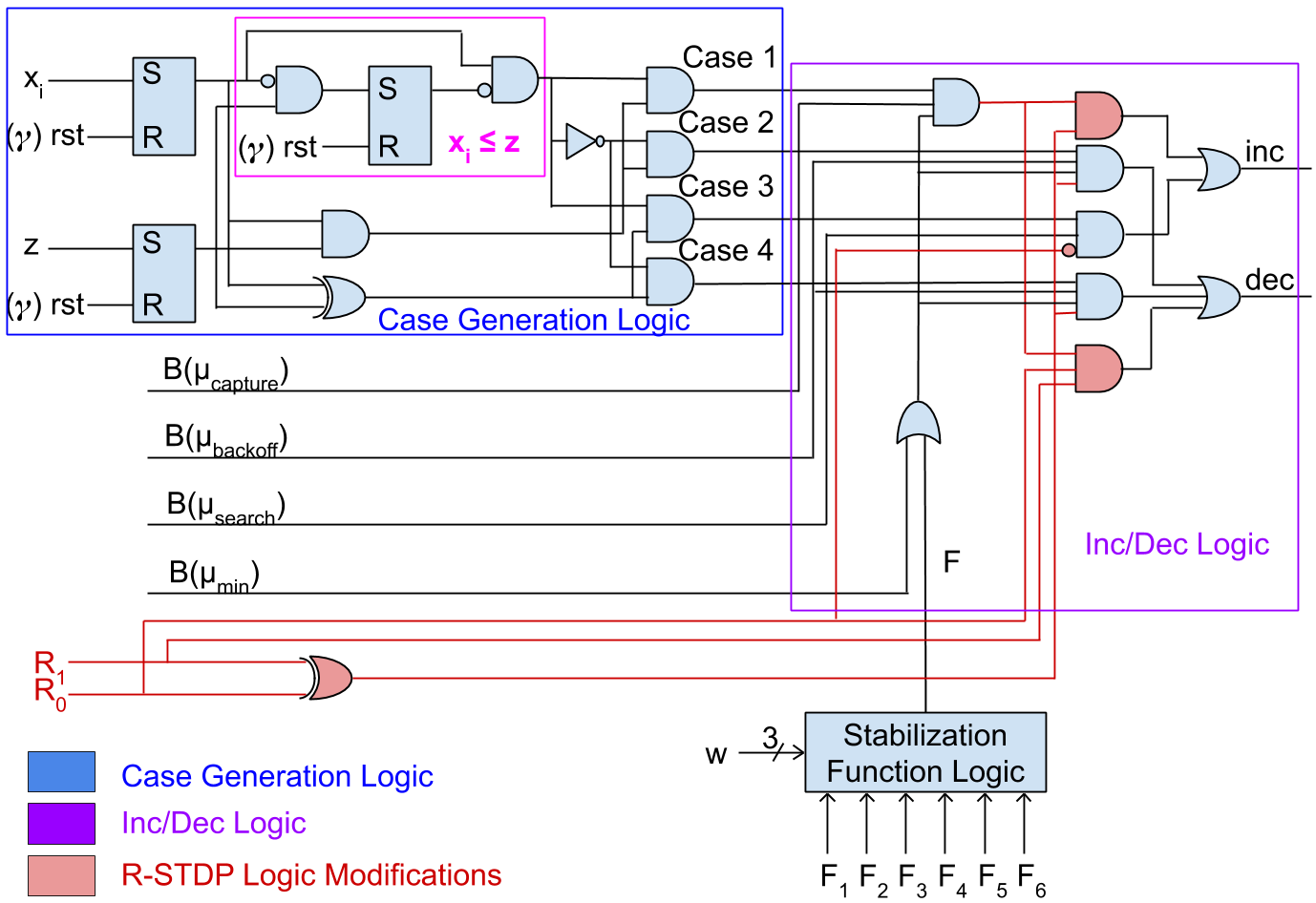}
        \caption{STDP and R-STDP Logic Implementation}
        \label{stdp_logic}
\end{figure}
The proposed STDP logic implementation is shown in Fig. \ref{stdp_logic}. It generates 2 control signals (increment/decrement) at the output that feed into the synaptic weight counters described in Fig. \ref{pt}. 
Note that STDP updates (and the associated resets) are performed at the end of a computational cycle (or onset of next gamma clock); inputs for the new computational cycle begin a unit clock cycle later.
The proposed STDP logic implementation can be partitioned into three components.

\subsubsection{Case Generation Logic}
The per-synapse case generation logic compares the synapse's input spiketime (\textit{x}\textsubscript{i}) with its post-synaptic neuron's output spiketime (\textit{z}) and generates 4 control signals corresponding to the 4 cases in Table \ref{stdp}. Case 5 is implicitly invoked when none of the other 4 cases is a 1.
The logic equations implemented for the 4 STDP cases are:

\begin{itemize*}
\item Case 1: $(x_i\leq z).(x_i).(z)$
\item Case 2: $(\overline{x_i\leq z}).(x_i).(z)$
\end{itemize*}

\begin{itemize*}
\item Case 3: $(x_i\leq z).(x_i\oplus z)$
\item Case 4: $(\overline{x_i\leq z}).(x_i\oplus z)$
\end{itemize*}

Note that ($(x_i\leq z)$) is implemented here using a much simpler \textit{temporal} comparator as opposed to a binary comparator. If \textit{z} arrives prior to \textit{x}, the output is 0; else \textit{x} is allowed to pass. 
%

\subsubsection{Stabilization Function Logic}
This logic selects 1 BRV from a set of finite BRVs generated by $F(w)$, based on the synaptic weight. For $w_{max}=7$, there are 6 non-zero BRVs to choose from. The output bit is generated by an 8-to-1 multiplexer controlled by 3-bit weight.

\subsubsection{Inc/Dec Logic}
The inc/dec logic assumes 4 BRV inputs from the LFSR network corresponding to the four STDP cases. The \textit{max} operation in Table \ref{stdp} is simply implemented by ‘OR’ing `F' with \textit{min} BRV input. The output of the stabilization logic is used along with the cases from case generation logic to generate \textit{inc} and \textit{dec} outputs. 

%
%
%
\subsection{Proposed R-STDP Implementation}
This subsection introduces a variation of the proposed STDP method capable of \textit{reinforcement learning} (R-STDP) \cite{mozafari2019bio} that uses an external \textit{reward} signal to drive its learning process in a desired direction.
It involves three forms of reinforcement:
\begin{itemize}[noitemsep, topsep=2pt]
    \item When the column's (non-null) output matches the desired action, \textit{reward} = `1'. It operates as per Table \ref{stdp}; except
    case 3 results in no synaptic weight update.
    \item When the column's (non-null) output does not match the desired action, \textit{reward} = `-1'. 
    Only Cases 1 and 3 are performed;    
    for Case 1, weight is actually decremented instead of incremented.
    \item When the column produces no output, i.e., no neuron spikes, \textit{reward} = `0' and only Case 3 operates.
\end{itemize}


In effect, desired behavior is reinforced and undesirable behavior is repressed using a single \textit{global} reward signal.
Note that R-STDP is still applied locally to each neuron and is typically deployed in the final layer of a TNN.
The logic modifications for R-STDP are minimal and straightforward as highlighted in Fig. \ref{stdp_logic}. \textit{reward} is a 2-bit signal (which encodes `-1', `0' and `1' as `11', `00' and `01' respectively). Unsupervised STDP is invoked when \textit{reward} is `10'. 
%
%

The implemented STDP/R-STDP learning rules are capable of performing extremely efficient online incremental learning (see Section \ref{online_eval}). To the best of our knowledge, such gate-level hardware-efficient implementations of STDP/R-STDP rules for TNNs have not been presented or published before.

\section{Column Implementation}
\label{column_sec}
A column is a fundamental functional unit in TNNs \cite{smith2020neuromorphic,smith2020temporal}, much like ALUs in von-Neumann computers.
As shown in Fig. \ref{fig3b_new}, a \textit{p}$\times$\textit{q} column contains \textit{q} excitatory neurons and a synaptic crossbar connecting the \textit{p} inputs to the \textit{q} neurons via \textit{p}$\times$\textit{q} synapses. A column supports unsupervised learning via STDP or supervised learning via R-STDP, followed by WTA lateral inhibition to assist in weight convergence. A single column supported by STDP/R-STDP and WTA becomes a fully operational TNN, capable of performing online continual learning and inferencing. Columns can be used to create larger TNNs by stacking multiple columns to form a multi-column layer, and by cascading multiple layers into a multi-layer TNN. 
\begin{figure}[t]
        \centering
        \includegraphics[width=2.6in]{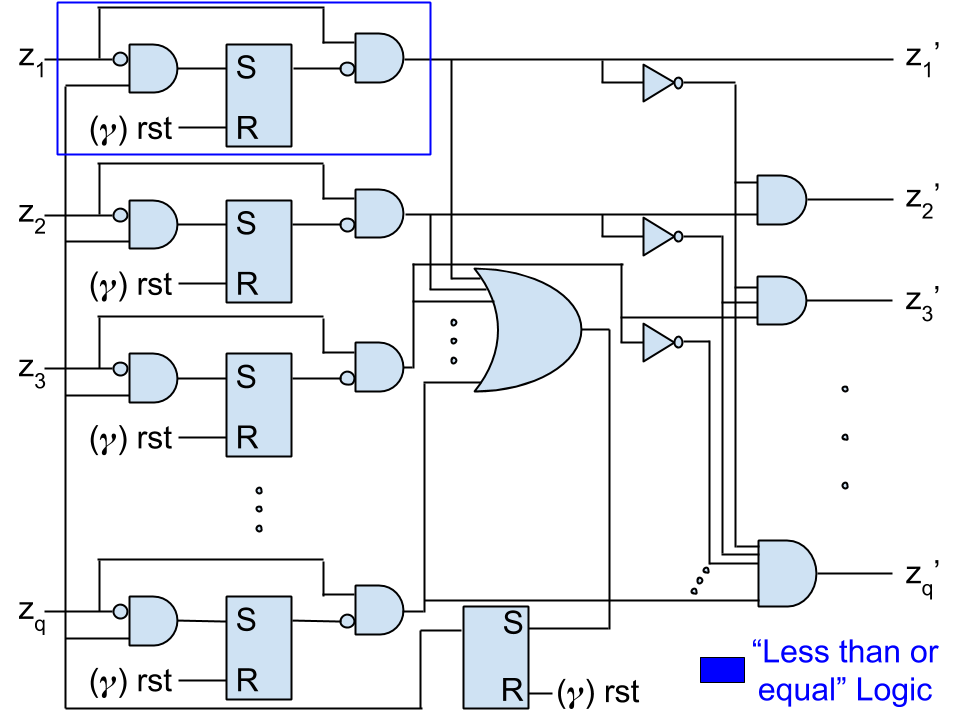}
        \caption{WTA Inhibition for a Column of \textit{q} Neurons}
        \label{li}
\end{figure}

Winner-take-all (WTA) inhibition is a distinctive feature of a column that selects the first spiking neuron and allows its output spike to pass through intact, while nullifying other neurons' outputs. 
Fig. \ref{li} shows the logic diagram for 1-WTA inhibition across \textit{q} neurons in a column.
The inhibition operation is performed by a latch-based less-than-or-equal temporal comparison unit.
The first spike is found through a large ‘OR’ gate, or a tree of small OR gates, (performing a temporal 'min' function) and is fed back through a latch which holds the signal at 1 until the next gamma cycle. Any input pulse coming to the latch after this signal is blocked, so only the first spikes are passed.
Tie breaking is implemented by selecting the spiking neuron with the lowest index. 

\section{Microarchitecture Framework Evaluation}
\label{uarch_eval}

Scalable neuron and column designs are implemented in System Verilog; synthesis results are generated using open-source 45nm Nangate standard cell library \cite{knudsen2008nangate} and Synopsys tools. Hardware design is evaluated in terms of area (A), critical path delay (D), computation time (T) and power (P). 
T is the time taken to process one input (one \textit{gamma} cycle).
\subsection{Gate-Level Characteristic Scaling Equations}
We derive characteristic scaling equations (Table \ref{gateeval}) for A, D (neuron), T (column) and P based on gate count ('AND' equivalents) and number of signal transitions, parameterized in terms of number of neurons (\textit{q}) and number of synapses per neuron (\textit{p}). The procedure is as follows: 1) Gate count is used as a surrogate for area and static power. 2) Number of gates in the critical path is used for D; T is derived using the \textit{gamma} period, $T = 15*D$. 3) Number of gate transitions is used for dynamic power. These equations can serve as a powerful tool for design space exploration, as they can help estimate the hardware complexity of arbitrary TNN designs. 

From our gate-level analysis for a single neuron, synapses (including STDP) constitute almost 90\% (50\% synaptic FSM and 40\% STDP logic) of the entire neuron complexity while the neuron body accounts for the remaining 10\%. In a single column, neurons constitute almost the entirety of column complexity; WTA incurs negligible cost (less than 1\%). 

\renewcommand{\tabcolsep}{2pt}
\begin{scriptsize}
\begin{table}[t]
  \centering
  \caption{Characteristic scaling equations for A, D/T and P for a neuron with \textit{p} synapses and a \textit{p}$\times$\textit{q} column.}
  \resizebox{9cm}{1.111cm}{
  \begin{tabular}{|c|c|c|}
    \hline
    \textbf{Metrics} & \textbf{Neuron} & \textbf{Column}\\
    \hline
    \hline
    A & $102p+8log_2p+36$ & $102pq+8qlog_2p+44q+q^2$ \\
    \hline
    D / T & $6log_2p+4$ & $90log_2p+60$ \\
    \hline
    P\textsubscript{static} & $102p+8log_2p+36$ & $102pq+8qlog_2p+44q+q^2$ \\
    \hline
    P\textsubscript{dynamic} & $204p+185log_2p+241$ & $204pq+185qlog_2p+257q+2q^2$ \\
    \hline
  \end{tabular}
  }
  \label{gateeval}
\end{table}
\end{scriptsize}
\renewcommand{\tabcolsep}{6pt}
\begin{scriptsize}
\begin{table}[t]
  \centering
  \caption{A, T and P (in 45 nm CMOS) for three column sizes of 64$\times$8, 128$\times$10, 1024$\times$16, with STDP and R-STDP.}
  \scalebox{1}{
  \begin{tabular}{|c|c|r|c|c|c|}
    \hline
    & \textbf{Synapses x} & \textbf{Gate} & \textbf{Area} & \textbf{Comp.} & \textbf{Power}\\
    & \textbf{Neurons} & \textbf{Count} & [mm\textsuperscript{2}] & \textbf{Time} [ns] & [mW]\\
    \hline
    \hline
    \multirow{3}{*}{STDP} & 64 $\times$ 8 & 51,824 & 0.05 & 28.95 & 0.25 \\
    \cline{2-6}
    & 128 $\times$ 10 & 128,658 & 0.13 & 32.40 & 0.62 \\
    \cline{2-6}
    & 1024 $\times$ 16 & 1,639,020 & 1.65 & 42.30 & 7.96 \\
    \hline
    \multirow{3}{*}{R-STDP} & 64 $\times$ 8 & 54,384 & 0.05 & 28.95 & 0.26 \\
    \cline{2-6}
    & 128 $\times$ 10 & 135,058 & 0.14 & 32.40 & 0.65 \\
    \cline{2-6}
    & 1024 $\times$ 16 & 1,720,940 & 1.75 & 42.30 & 8.36 \\
    \hline
  \end{tabular}
  }
  \label{cresults}
\end{table}
\end{scriptsize}
\renewcommand{\tabcolsep}{2pt}

\subsection{Post-synthesis Evaluation of Column Designs}
Area, power and critical path delay are obtained directly from Design Compiler, and computation time is derived as earlier. 
We use the low power process corner for synthesis with operating frequency of 100 kHz and voltage of 0.95 V. 

Table \ref{cresults} presents 45 nm post-synthesis results for three column configurations for STDP and R-STDP learning rules: 1) a small 64$\times$8 column; 2) a medium 128$\times$10 column; and 3) a large 1024$\times$16 column. The \textit{gamma} cycle for the large 1024$\times$16 column with around 1.7M gates is 42.3 ns (23.64 MHz). It
has an area and power footprint of 1.65 mm\textsuperscript{2} and 7.96 mW with STDP in 45nm, less than 1\% of the area and power budget of typical mobile SoCs. 
Note that the overhead for R-STDP is minimal; it increases die area and power by only 5\% relative to STDP while adding supervision to learning.

\section{Online Incremental Learning}
\label{online_eval}

\begin{figure}[t]
    \centering
    \begin{subfigure}{0.2\textwidth}
        \centering
        \includegraphics[width=\textwidth]{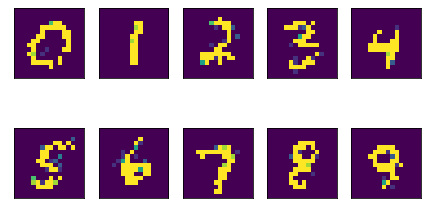}
        \caption{Trained for Digits 0 - 9}
        \label{fig_14a}
    \end{subfigure}%
    \hspace{15pt}
    \begin{subfigure}{0.2\textwidth}
        \centering
        \includegraphics[width=\textwidth]{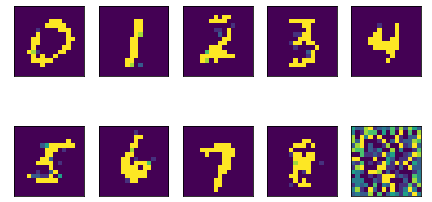}
        \caption{Trained for Digits 0 - 8}
        \label{fig_14b}
    \end{subfigure}
    \label{w_num}
    \caption{Synaptic weight matrices converge to image centers resembling MNIST digits in 10,000 samples.}
\end{figure}

\begin{figure}[t]
        \centering
        \includegraphics[width=3.43in]{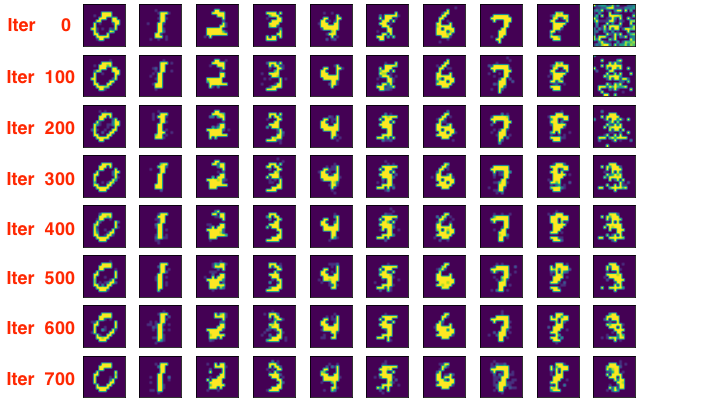}
        \caption{Online Incremental Learning: STDP learns a previously unseen input number '9' within 500 examples.}
        \label{inc_learn}
\end{figure}

In contrast to the typical epoch-based training with global back-propagation, STDP is an online local learning method that processes inputs in a streaming manner targeting online real-time applications. This section uses a subset of MNIST, with images resized from 28$\times$28 to 16$\times$16, to illustrate online incremental learning for TNNs. Because our focus is on online learning, the standard MNIST benchmark protocol for offline training/testing does not apply.
From our experiments, using just a single (256$\times$10) column and resized MNIST, several interesting capabilities of TNNs can be observed.
\begin{enumerate}[noitemsep, topsep=2pt]
    \item \textit{Online Classification via Centroid Formation}: Fig. \ref{fig_14a} shows the synaptic weights converged to the 10 class centroids via R-STDP, which resemble the corresponding digits. This shows the efficacy of R-STDP in driving the weights towards class centroids.
    \item \textit{Fast Training Convergence}: The synaptic weights in Fig. \ref{fig_14a} and Fig. \ref{fig_14b} converged after approximately 10,000 training samples, which implies that TNNs can learn very quickly and can generalize from  small datasets.
    \item \textit{Online Incremental Learning}: In this experiment, supervised R-STDP training is first performed with only 9 classes (0 to 8) by hiding the digit '9', resulting in the converged weights shown in Fig. \ref{fig_14b}. Then the digit '9' is introduced in the input sequence without labels to illustrate the ability to dynamically learn a previously unseen class 
    in an unsupervised fashion. As shown in Fig. \ref{inc_learn}, the synaptic weights converge to the digit '9' after only about 500 testing samples via STDP. 
\end{enumerate}
Thus, 
online incremental learning enables a TNN to adapt to new input data not seen before during the original (offline) training. Continual learning allows a TNN to keep learning and improving its performance concurrently with inference.

\section{Concluding Remarks}
Previous works in \cite{smith2020temporal,smith2020neuromorphic,chaudhari2021unsupervised,kheradpisheh2016bio,mozafari2019bio, dong2018unsupervised} have shown that TNNs can achieve online brain-like sensory processing and learning for vision and time-series applications. This work proposes a microarchitecture framework for directly implementing arbitrary TNNs using the building blocks: neurons, columns and online STDP/R-STDP learning. This work demonstrates the hardware implementation feasibility of TNNs using off-the-shelf CMOS technology and design tools, and represents an initial step in a promising direction for future research. The implementation results in this work should be viewed as a first opportunistic attempt, using existing design methods and tools. There are promising innovations, including custom macro cells and novel synthesis tools, that can be developed to further enhance the proposed design framework.

\bibliographystyle{IEEEtranS}
\bibliography{refs}

\end{document}